\documentclass[twocolumn,prl,tightenlines,superscriptaddress,showpacs]{revtex4-1}

\usepackage{amsmath}
\usepackage{amssymb,amsfonts,latexsym}
\usepackage{bm}
\usepackage[mathcal]{euscript}
\usepackage{graphicx}
\usepackage{epsfig}
\usepackage{color}
\usepackage{xfrac}
\usepackage{mathrsfs}

\usepackage{hyperref}
\usepackage{url}

\begin{document}

\title{Long-Range Nematic Order in Two-Dimensional Active Matter}

\author{Beno\^{\i}t Mahault}
\affiliation{Max Planck Institute for Dynamics and Self-Organization (MPIDS), 37077 G\"ottingen, Germany}

\author{Hugues Chat\'{e}}
\affiliation{Service de Physique de l'Etat Condens\'e, CEA, CNRS Universit\'e Paris-Saclay, CEA-Saclay, 91191 Gif-sur-Yvette, France}
\affiliation{Computational Science Research Center, Beijing 100193, China}

\date{\today}

\begin{abstract}
Working in two space dimensions, we show that the orientational order emerging from self-propelled polar particles aligning nematically
 is quasi-long-ranged beyond $\ell_{\rm r}$, the scale associated to induced velocity reversals,
which is typically extremely large and often cannot even be measured.
Below $\ell_{\rm r}$, nematic order is long-range.
We construct and study a hydrodynamic theory for this de facto phase 
and show that its structure and symmetries differ from conventional descriptions of active nematics. 
We check numerically our theoretical predictions, in particular the presence of 
$\pi$-symmetric propagative sound modes, and
provide estimates of all scaling exponents governing long-range space-time correlations.
\end{abstract}

\maketitle

Studies of active matter continue to flourish, exploring more and more complex situations in an increasingly quantitative manner \footnote{For recent experimental works, see, e.g. 
\cite{soni2019odd,li2019data,sugi2019Celegans,bain2019dynamic,duclos2020topological,strubing2020wrinkling,deblais2020phase,yamauchi2020chiralitydriven,sciortino2021pattern,rajabi2021directional,zhang2021spatiotemporal}.}. 
Evidence accumulates showing that active matter exhibits collective properties impossible in thermal equilibrium or even in driven systems \footnote{For recent theoretical works, see, e.g. 
\cite{tjhung2018cluster,baek2018generic,mahault2018self,souslov2019topological,pietzonka2019autonomous,scheibner2020non-Hermitian,saha2020scalar,dadhichi2020nonmutual,zakine2020surface,you2020nonreciprocity,banerjee2021active,denk2020pattern-induced,meng2021magnetic,reichhardt2021active}
.}. 
In spite of all this progress, important fundamental questions remain open.
A long-standing such issue 
is whether true long-range nematic order can emerge in two space dimensions (2D). 

Whereas it is now well known, notably thanks to the seminal work by Toner and Tu,
that long-range polar order can arise in 2D active systems \cite{toner1995long,toner1998flocks,tu1998sound,toner2012reanalysis,toner2012birth,mahault2019TT,Chen_Lee_Toner_2016,maitra2020swimmer,chate2020dry}, the debate has remained opened for active nematics: 
On the one hand, theoretical results conclude that nematic order can at best be quasi-long-range \footnote{The only exception we are aware of is \cite{maitra2021stable} but the coupling of the 2D active layer to a surrounding 3D
fluid is essential to insure long-range order.},
as in equilibrium, albeit with important differences
\cite{simha2002hydrodynamic,ramaswamy2003active,toner2005hydrodynamics,mishra2010dynamic,marchetti2013hydrodynamics,shankar2018low,maitra2018nonequilibrium}. On the other hand numerical and experimental results obtained on self-propelled particles without spontaneous velocity reversals yielded convincing data demonstrating true long-range nematic order over a large range of scales \cite{ginelli2010large,nishiguchi2017long}.

In this Letter, we study 2D dry dilute active nematics 
---the framework in which the question of the asymptotic nature of nematic order was mostly discussed--- using numerical simulations and theory. We show that the homogeneous ordered phase of a Vicsek-style model of polar self-propelled particles aligning nematically actually displays true long-range nematic order only up to  $\ell_{\rm r}$, the scale associated to typical time between velocity reversals {\it induced} by collisions and noise. Beyond $\ell_{\rm r}$, global nematic order decays algebraically with system size, in agreement with general theoretical arguments. 
However $\ell_{\rm r}$ can easily take astronomically large values 
such that there exists a region of parameter space in which only true long-range nematic order can be observed. 
We derive a hydrodynamic theory for this regime and show that it possesses a structure and symmetries different from those of standard active nematics. Our analysis of this field theory predicts $\pi$-symmetric sound modes 
and the scaling form of space-time fluctuations. 
Finally, numerical results confirm the theory and allow us to estimate all scaling exponents.

We use the Vicsek-style model of polar particles with nematic alignment first introduced in \cite{ginelli2010large}. 
Particles $i=1,\ldots,N$ evolve
at discrete timesteps with constant speed $v_0$ in square domains of linear size $L$ with periodic boundary conditions, interacting with neighbors within unit distance. Their positions ${\bf r}_i$ and unit-length orientations ${\bf e}_i= {\bf e}(\theta_i)$ obey:
\begin{subequations}
\begin{eqnarray}
	{\bf r}_i^{t+1} &=& {\bf r}_i^t + v_0 {\bf e}_i^{t +1} , \label{eqvm3}\\
	{\bf e}_i^{t+1} &=& \left( {\mathcal R}_\eta \circ \vartheta \right) 
	\langle {\rm sign}[{\bf e}_i^t\cdot {\bf e}_j^t  ] {\bf e}_j^t\rangle_{j\sim i} ,
	\label{eqvm4}
\end{eqnarray}
\label{eqvm}
\end{subequations}
where $\vartheta$ normalizes vectors ($\vartheta({\bf u})={\bf u}/\|{\bf u}\|$),
and ${\mathcal R}_\eta$ rotates them by a random angle drawn from a uniform distribution in
$[-\pi\eta,\pi\eta]$, independently for every particle at every timestep.
The two main parameters are the global density $\bar{\rho}=N/L^2$ and the noise strength $\eta$. 
The phase diagram in the $(\bar{\rho},\eta)$ plane is typical of Vicsek-style models \cite{chate2020dry}.
All results presented below were obtained with $v_0=0.5$ and $\bar{\rho}=2$.

\begin{figure}[t!]
\includegraphics[width=\columnwidth]{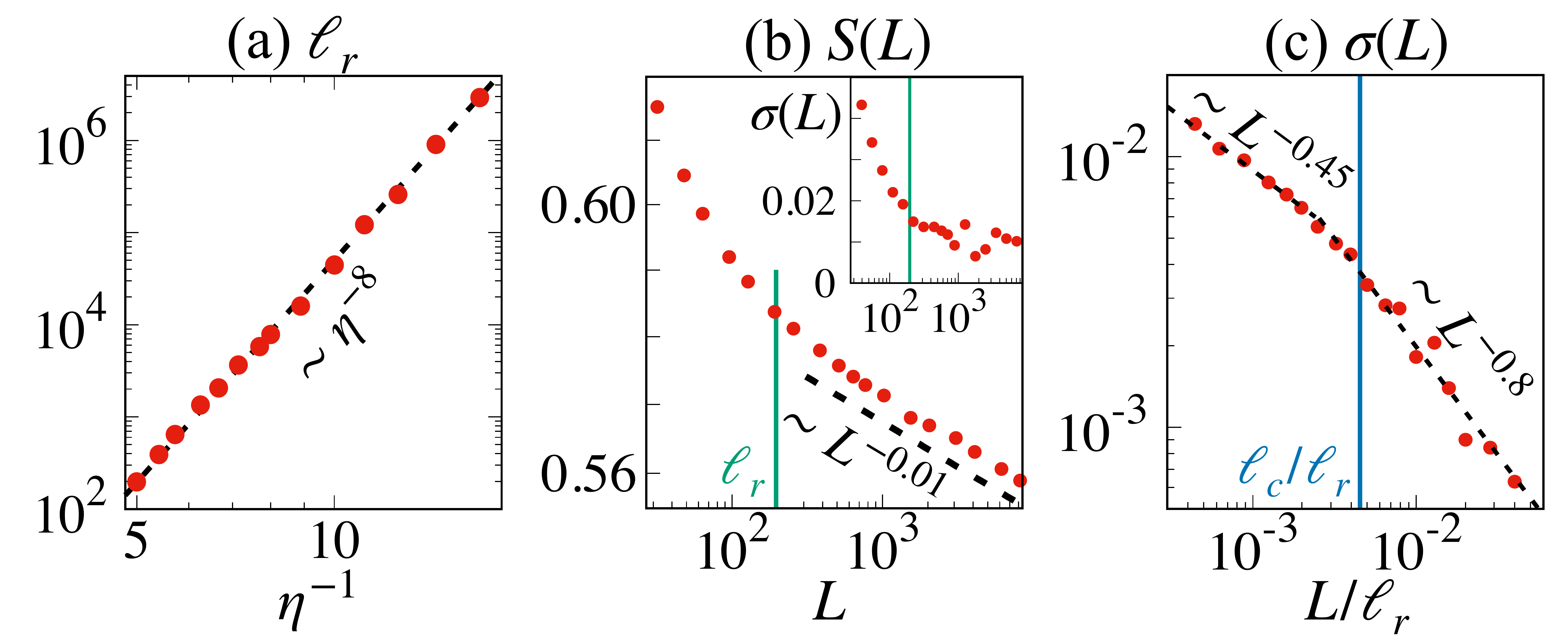}
\caption{Vicsek-style model \eqref{eqvm} ($v_0=0.5$, $\bar{\rho}=2$).
(a) Variation of $\ell_r$ with $\eta$.
(b) Global nematic order $S$ vs linear system size $L$ (in log scales); 
for $L<\ell_{\rm r}\simeq200$, $S$ decreases slower than a powerlaw,
while a slow algebraic decay is observed for $L \gg \ell_r$ ($\eta=0.2$);
inset: local slope $\sigma(L)$ showing a plateau for $L \gg \ell_r$.
(c) $\sigma(L)$ vs $L/\ell_r$ in the long-range ordered regime
($\eta=0.1$, for which $\ell_{\rm r}\simeq50000$);
Note the crossover scale $\ell_{\rm c}\simeq 200$ separating two scaling regimes.
}
\label{fig1}
\end{figure}

We focus on the homogeneous nematic liquid that exists for $\eta\lesssim 0.21$,
where the global nematic order parameter $S= \langle|\langle e^{i2\theta_k^t}\rangle_{k}|\rangle_t$ takes ${\cal O}(1)$ values.
In this state, particles can be split into two `polar' subpopulations according to which of the two opposite directions
defined by the nematic order their orientation is closest. The nematic interaction in Eq.~\eqref{eqvm4} aligns
particles belonging to the same population and anti-aligns particles belonging to opposite populations, so that
particles mostly stay in the same population. Nevertheless, under the action of interactions and noise, they
can eventually turn enough that they join the other population.
It was shown in \cite{ginelli2010large} that the distance traveled between such reversals is distributed exponentially 
with a characteristic length $\ell_{\rm r}$ independent of system size. In Fig.~\ref{fig1}(a), we show that $\ell_{\rm r}$ grows
very fast when the noise strength $\eta$ decreases. A good fit of our data is that $\ell_{\rm r}\sim \eta^{-8}$.

In \cite{ginelli2010large}, the global nematic order parameter $S$ was found to decrease slower than a power of $L$ 
and consistent with an algebraic decay to a finite asymptotic value ($S(L)-S(\infty)\sim L^{-\varpi}$). 
These results led to conclude to true long-range nematic order, but they were obtained 
on a range of system sizes barely encompassing $\ell_{\rm r}$. Here, choosing a noise strength such that
$\ell_{\rm r}$ is not too large, we find that for $L>\ell_{\rm r}$, $S$ decays like a small power of $L$,
in departure from the $L<\ell_{\rm r}$ behavior (Fig.~\ref{fig1}(b)). 
Asymptotically, nematic order is only quasi-long-range, in agreement with standard theories \cite{shankar2018low}.

Nevertheless, in most of the homogeneous nematic phase, $\ell_{\rm r}$ is so large that only the 
$L<\ell_{\rm r}$ regime is accessible and it is thus important to study it {\it per se}.
Working in this regime,
we confirm that nematic order is fully long-range; moreover,
the scaling of the local slope $\sigma(L) \equiv -{\rm d}\ln(S)/{\rm d}\ln(L) \sim L^{-\varpi}$ 
allows to identify an internal crossover scale $\ell_{\rm c}$ separating two regimes with
different values of $\varpi$ (Fig.~\ref{fig1}(c)).

We now present a theory of the long-range-ordered nematics present on scales 
much smaller than $\ell_{\rm r}$. Full details of calculations are given in \cite{SUPP}.
Our approach is not a perturbative version of active nematics:
We directly consider two populations, $R$ and $L$, of polar active particles 
with speed $v_0$ aligning their velocity with neighbors if those
belong to the same population, and anti-align it otherwise. 
This is {\it not} equivalent to usual nematic alignment:
two particles of the same population will align even if their relative angle is obtuse,
and they will anti-align if they belong to different populations, irrespective of their angle.
We further assume that the populations exchange members randomly at rate $1/\tau_{\rm r}\simeq\ell_{\rm r}/v_0$.
We first write Boltzmann equations ruling the 
evolution of the one-body probability density functions $f_L({\bf r},\theta,t)$ and $f_R({\bf r},\theta,t)$:
\begin{equation} \label{eq_Boltzmann}
\partial_t f_{L} + {\bf v}(\theta)\cdot \nabla f_{L} = 
\tfrac{1}{\tau_{\rm r}}(f_{R} - f_{L}) + I_{\rm sd}[f_{L}] + I_{\rm co}[f_{L},f_{R}],
\end{equation}
and the equation governing $f_R$ is given by swapping the $L$ and $R$ subscripts.
In \eqref{eq_Boltzmann}, ${\bf v}(\theta)=v_0{\bf e}(\theta)$ is the velocity of particles 
with orientation $\theta$, 
whereas the integrals $I_{\rm sd}$ and $I_{\rm co}$, given in \cite{SUPP}, describe the effects of angular self-diffusion and collisions.

Introducing the more convenient $f=f_R+f_L$ and $g=f_R-f_L$, 
expanding $f$ and $g$
in Fourier series of $\theta$ 
(e.g. $f({\bf r},\theta,t)=\tfrac{1}{2\pi}\sum_{k=-\infty}^{+\infty} f_k({\bf r},t)e^{-ik\theta}$),
the Boltzmann equations are de-dimensionalized and transformed into a hierarchy of partial differential equations 
for the $f_k$ and $g_k$ fields.
As shown in \cite{SUPP}, a linear stability analysis of the disordered solution $\rho \equiv f_0 = \bar{\rho}$ (the total density), $f_{k> 0} = g_k = 0$ 
reveals that it is unstable to $g_1$ perturbations at large density and/or weak noise.
The field $g_1$ is thus responsible for the onset of orientational order. 
Note that $g_1$ measures polar order within each population, {\it i.e.} is a proxy for global nematic order. 
The equations for $\rho$ and $g_0$ read
\begin{subequations}
\label{eq_O4_hydro}
\begin{align}
\partial_t \rho & = - {\rm Re}[ \triangledown^* f_1] \,, \label{eq_rho}\\ 
\partial_t g_0 & = -2 \tau_r^{-1} g_0 - {\rm Re}[ \triangledown^* g_1 ] \,, \label{eq_g0}
\end{align}
where $\triangledown \equiv \partial_x + i\partial_y$ denotes the complex gradient.

Following the Boltzmann-Ginzburg-Landau approach
\cite{bertin2013mesoscopic,peshkov2014boltzmann,DADAM_LesHouches,chate2020dry}, 
one can build step by step a scaling ansatz
using a small parameter $\varepsilon$ marking the magnitude of order near onset ($|g_1|\sim\varepsilon$).
As detailed in \cite{SUPP}, this leads to:
$ |g_{k \ge 1}| \sim \varepsilon^k$, $|f_{k > 1}| \sim \varepsilon^k$, and 
$\partial_t \sim \triangledown \sim \varepsilon$~\footnote{Note that this propagative ansatz is at odds with
the diffusive one usually at play in active nematics}.
In addition, considering Eqs.~(\ref{eq_rho},\ref{eq_g0}),
one completes the scaling ansatz by
$|g_0| \sim \varepsilon \,, |\delta \rho| \sim |f_{1}| \sim \varepsilon^2$. 
Truncating and closing the Boltzmann hierarchy at order $\varepsilon^4$
yields hydrodynamic equations for $f_1$ and $g_1$:
\begin{align}
\partial_t f_1  = & -\tfrac{1}{2} \triangledown \rho + \left( \mu[\rho] - \zeta |g_1|^2 \right)  f_1
+ D_{\!f} \Delta f_1 \nonumber \\
& + \left( \alpha[g_0] -\chi_1 g_0 |g_1|^2  - \chi_2 f_1^* g_1 \right) g_1 
 + D_g g_0 \Delta g_1 \nonumber \\
\label{eq_hydro_f1_O4} &  + \kappa_1[\rho] \triangledown^* g_1^2 + \kappa_2[\rho] g_1^* \triangledown g_1 + \kappa_3 (\triangledown^* g_0)(\triangledown g_1)  \\
\partial_t g_1 = & -\tfrac{1}{2} \triangledown g_0  + \left( \nu[\rho] - \Gamma[\rho] |g_1|^2 \right) g_1  + \Omega[\rho] \Delta g_1 \nonumber \\
&- \sigma g_1^2\triangledown^* g_0 + \beta[g_0] f_1 + \lambda_1 g_0 \triangledown^* g_1^2 + \lambda_2 g_0 g_1^* \triangledown g_1\nonumber \\
\label{eq_hydro_g1_O4} &  + \lambda_3 g_1^* \triangledown f_1  + \lambda_4 \triangledown^*(g_1 f_1) + \lambda_5 f_1^* \triangledown g_1
\end{align}
\end{subequations}
where all coefficients 
depend on the particle-level parameters $\bar{\rho}$, $\eta$ and $\tau_r$.
(see \cite{SUPP} for their explicit expressions), 
and local dependencies on $\rho$ and $g_0$ are indicated. 

Eqs.~\eqref{eq_O4_hydro},
are structurally different from hydrodynamic theories written for active nematics.
The $2\pi$-symmetry of the interaction between our polar particles makes
the pairs of equations for $(\rho,f_1)$ and $(g_0,g_1)$ resemble two coupled Toner-Tu (TT) systems.
Both $\rho$ and $g_0$ are advected by the corresponding order fields $f_1$ and $g_1$, 
which are {\it not} $\pi$-symmetric. 
Discarding the couplings to $\rho$ and $f_1$, Eqs.~\eqref{eq_g0} and~\eqref{eq_hydro_g1_O4} are {\it almost} like the TT equations in the limit $\tau_r \to \infty$.
They however miss terms $\sim g_0 g_1$ and $\sim g_1 \triangledown g_1$ 
that are forbidden by the $R \leftrightarrow L$ symmetry of the problem,
which imposes the equations to be invariant under $g \leftrightarrow -g$.

Eqs.~\eqref{eq_O4_hydro}, 
even if formally derived at the onset of order, reflect the symmetries of the deeply ordered phase. 
We now focus on fluctuations in that phase, {\it i.e.} around the homogeneous ordered solution $\rho = \bar{\rho}$, $g_0 = f_1 = 0$, $g_1 = \bar{g} \equiv \sqrt{ \nu[\bar{\rho}] / \Gamma[\bar{\rho}] }$
that exists when $\nu[\bar{\rho}]>0$\footnote{Choosing $\bar{g}$ real assumes, with loss of generality, that order is along $x$.}. 
Linearizing Eqs.~\eqref{eq_O4_hydro} 
around this solution, 
separating parallel ($\|$) and transverse ($\perp$) components, we obtain a system of 6 equations
governing small perturbations $\delta\rho$, $\delta g_0$, $\delta f_\|$, $\delta f_\perp$, $\delta g_\|$,
and $\delta g_\perp$.
We find that $\delta \rho$ and $\delta g_\perp$ are hydrodynamic modes, 
while $\delta f_\|$, $\delta f_\perp$, and $\delta g_\|$ decay rapidly.
Since $\tau_r$ can take arbitrary large values, 
we also consider $\delta g_0$ as hydrodynamic.
Enslaving the fast modes, we obtain the following linear system:
\begin{subequations}
\label{eq_lin_closed}
\begin{align}
\label{eq_delta_rho_closed}
\partial_t \delta\rho =& (D_{\rho\|} \partial^2_{\|\|} + D_{\rho\perp}\partial^2_{\perp\perp})\delta \rho\nonumber \\
 &- \lambda_0 \partial_\| \delta g_0 + D_{\rho g} \partial^2_{\| \perp} \delta g_\perp
+ \eta_1 \partial_{\| t}^2 \delta g_0  , \\
\label{eq_delta_g0_closed}
\partial_t \delta g_0 =& ( D_{0\|} \partial^2_{\|\|} + D_{0\perp} \partial^2_{\perp\perp} - 2 \tau_r^{-1}) \delta g_0\nonumber \\
& - \kappa_0 \partial_\| \delta \rho - v_0 \partial_\perp \delta g_\perp 
+ \eta_2 \partial_{\| t}^2 \delta \rho , \\
\label{eq_delta_g1_closed}
\partial_t \delta g_\perp =& ( D_\| \partial^2_{\|\|} + D_\perp \partial^2_{\perp\perp}) \delta g_\perp \nonumber\\
& + \gamma \partial^2_{\| \perp}\delta\rho - \alpha_0 \partial_\perp \delta g_0 
+ \eta_3 \partial_{\perp t}^2 \delta g_0 ,
\end{align}
\end{subequations}
where we split the complex gradient into $\triangledown = \partial_\| + i\partial_\perp$, and all the (bare)
coefficients are given in \cite{SUPP}\footnote{We included in Eq.~\eqref{eq_delta_g0_closed} the term $D_{0\perp} \partial^2_{\perp\perp}\delta g_0$, allowed by symmetries, even though it does not appear in the derivation.}.

We first note that in the small $\tau_{\rm r}$ limit, 
such that particles reverse their orientation many times on the scale at which we observe fluctuations,
$\delta g_0$ is non-hydrodynamic (cf. Eq.~\eqref{eq_delta_g0_closed}). 
Eqs.~\eqref{eq_lin_closed} then reduce to those of an homogeneous active nematic 
(with $\delta g_\perp$ playing the role of the transverse fluctuations of nematic order, see \cite{SUPP,shankar2018low}).
 
In the $\tau_{\rm r}\to\infty$ limit of main interest here, on the other hand, we neglect the term $2\tau_r^{-1} \delta g_0$ in Eq.~\eqref{eq_delta_g0_closed}. 
To compute space and time correlation functions of the three hydrodynamic fields 
$\delta\rho$, $\delta g_0$ and $\delta g_\perp$,
we equip Eqs.~\eqref{eq_lin_closed} with additive, uncorrelated, zero-mean noise terms. 
For Eq.~\eqref{eq_delta_rho_closed}, governing density fluctuations, this noise is conserved 
and we write it $\partial_\| h_{\rho\|} + \partial_\perp h_{\rho\perp}$.
Writing the (stochastic) Eqs.~\eqref{eq_lin_closed} in Fourier space, taking the long wavelength, low frequency limit
$q,\omega\to 0$, rather tedious but standard calculations detailed in \cite{SUPP} lead to:
\begin{widetext}
\begin{subequations}
\label{eq_space_time_cfs}
\begin{align}
\left \langle \left|\delta\hat{\rho}(\omega,{\bm q}) \right|^2 \right\rangle& \underset{\omega,q \to 0}{\simeq} \;
{\cal D}(\omega,{\bm q})^{-1} \left[
(q_\|^2 \Delta_{\rho\|}+q_\perp^2 \Delta_{\rho\perp})(\omega^2- v_0\alpha_0q_\perp^2)^2+ \Delta_0 q^2_\| \lambda_0^2 \omega^2 + \Delta_\perp (v_0 \lambda_0)^2 q^2_\| q^2_\perp\right] \,, \\\label{eq_space_time_cfs_rho}
\left \langle \left|\delta\hat{g}_0(\omega,{\bm q}) \right|^2 \right\rangle& \underset{\omega,q \to 0}{\simeq} \;{\cal D}(\omega,{\bm q})^{-1} \left[(q_\|^2 \Delta_{\rho\|}+q_\perp^2 \Delta_{\rho\perp}) \kappa_0^2 q^2_\|\omega^2 + \Delta_0 \omega^4 + \Delta_\perp v^2_0 q_\perp^2 \omega^2\right] \,, \\
\label{eq_space_time_cfs_gperp}
\left \langle \left|\delta\hat{g}_\perp(\omega,{\bm q}) \right|^2 \right\rangle& \underset{\omega,q \to 0}{\simeq}\; {\cal D}(\omega,{\bm q})^{-1} \left[(q_\|^2 \Delta_{\rho\|}+q_\perp^2 \Delta_{\rho\perp})(\alpha_0\kappa_0)^2 q_\|^2q_\perp^2 + \Delta_0 \alpha_0^2 q_\perp^2 \omega^2 + \Delta_\perp(\omega^2 - \kappa_0\lambda_0 q_\|^2)^2 \right] \,,
\end{align}
\end{subequations}
where $\Delta_{\rho\|}$ and $\Delta_{\rho\perp}$ are the amplitudes of the conserved $\rho$ noise, 
$\Delta_0$ and $\Delta_\perp$ those of the $g_0$ and $g_\perp$ noises, 
and 
\begin{equation}
\label{eq_denom}
{\cal D}(\omega,{\bm q}) \equiv  
  |\omega -  i \varepsilon_{\rm d}({\bm q})|^2 \times |\omega - c(\theta_{\bm q}) q + i \varepsilon_{\rm p}({\bm q})|^2 \times  |\omega + c(\theta_{\bm q}) q + i \varepsilon_{\rm p}({\bm q})|^2 .
\end{equation}
\end{widetext}
As shown in \cite{SUPP}, where their explicit forms are given, $\varepsilon_{\rm d,p}({\bm q})\sim q^2$, 
whereas the anisotropic speed is
\begin{equation}
\label{eq_c}
c(\theta_{\bm q})  = \sqrt{ \kappa_0 \lambda_0 \cos^2(\theta_{\bm q}) + v_0 \alpha_0\sin^2(\theta_{\bm q}) }, 
\end{equation}
where $\theta_{\bm q}$ denotes the angle between $\bm q$ and the mean order.
Eqs.~(\ref{eq_space_time_cfs},\ref{eq_denom},\ref{eq_c}) 
are fundamentally different from 
their counterparts in both active nematics and the TT class: 
at most orientations $\theta_{\bm q}$ correlations have a 
diffusive peak 
and two symmetric propagative peaks at $\omega=\pm c(\theta_{\bm q})q$.

Equal-time correlation functions are easily obtained by integrating Eqs.~\eqref{eq_space_time_cfs} over $\omega$.
They all diverge as $q^{-2}$ for most $\theta_{\bm q}$, which means that nematic order is only quasi-long-range
at this linear level, a situation similar to that of polar order in TT theory.
To resolve this marginal situation, one needs to study nonlinear hydrodynamics. 
We first repeat the calculations leading to Eqs.\eqref{eq_lin_closed} 
keeping the leading order nonlinearities (in fields and gradients). 
The structure of our theory shares similarities with the polar case.
We thus limit ourselves to terms of order 3 in fields and gradients \cite{toner2012reanalysis}.
After lengthy but straightforward manipulations 
(detailed in \cite{SUPP}), we obtain:
\begin{subequations}
\label{eq_nonlin_closed}
\begin{align}
\label{eq_nonlin_delta_rho_closed}
\!\!\!\partial_t \delta\rho &= {\cal L}_\rho + j_1 \partial_\| (\delta g_0\delta\rho)
 + j_2 \partial_\perp(\delta g_0\delta g_\perp), \\
\label{eq_nonlin_delta_g0_closed}
\!\!\!\partial_t \delta g_0 &= {\cal L}_{g_0}
+ c_1 \partial_\| \delta\rho^2 + c_2 \partial_\| \delta g_\perp^2 + c_3 \partial_\| \delta g_0^2 , \\
\label{eq_nonlin_delta_g1_closed}
\!\!\!\partial_t \delta g_\perp & = {\cal L}_{g_\perp}
+ w_1 \delta g_\perp\partial_\|\delta g_0 + w_2 \delta g_0\partial_\|\delta g_\perp\nonumber \\
&  + w_3 \delta g_0\partial_\perp\delta\rho + w_4 \delta \rho\partial_\perp\delta g_0
+ w_5 \delta g_\perp\partial_t \delta \rho ,
\end{align}
\end{subequations}
where ${\cal L}$ is the linear part (Eqs.~\eqref{eq_lin_closed}).
Introducing the scaling exponents via
$x_\perp \to b x_\perp$, $x_\| \to b^\xi x_\|$, $t \to b^z t$, $\delta g_\perp \to b^\chi \delta g_\perp$,
$\delta g_0 \to b^{\chi_0} \delta g_0$, $\delta \rho \to b^{\chi_\rho} \delta\rho$
and imposing a fixed point condition on Eqs.~\eqref{eq_nonlin_closed} considered valid in any dimension $d$
yields the following values of the exponents 
\begin{equation}
\label{lin_exps}
z = 2, \, \xi = 1, \, \chi = \chi_{0} = \chi_{\rho} = 1 - \tfrac{d}{2} \;\;\;{\rm (linear \; level)}.
\end{equation}
We thus have isotropic $(\xi = 1)$ diffusive $(z=2)$ scaling
with quasi-long-range order in $d=2$ $(\chi = 0)$ at the linear level, as for both active nematics and TT theory.

At the linear fixed point, 9 of the 10 nonlinear terms in Eqs.~\eqref{eq_nonlin_closed} scale like $b^{(4-d)/2}$,
i.e. are relevant in $d \le d_c \equiv 4$ (the exception is $\omega_5$).
This means that the linear theory breaks down in $d \le 4$, and that we should in principle embark on a complete renormalization group analysis to obtain exponent values.
We leave this challenging task 
for future studies.
Instead we rely on general considerations and formal similarities with TT theory to make predictions
that we test numerically.

Replacing the eigenfrequencies $\omega = c(\theta_{\bm q})q$ into 
\eqref{eq_delta_rho_closed} and \eqref{eq_delta_g0_closed} (expressed in Fourier space), 
we find at leading order 
$c(\theta_{\bm q})q \delta\hat{\rho} \sim q_\| \delta\hat{g}_0$ and 
$c(\theta_{\bm q})q \delta\hat{g}_0 \sim q_\| \delta\hat{\rho} + q_\perp\delta\hat{g}_\perp$.
Therefore, taking $\theta_{\bm q} = \pi/2$ we get that $|\delta\hat{g}_0|\sim|\delta\hat{g}_\perp|$, such that $\chi_0 = \chi$, 
while for any orientation of $\bm q$ not purely longitudinal or transverse we have $|\delta\hat{\rho}|\sim|\delta\hat{g}_0|$, which implies $\chi_\rho = \chi_0$.
It is thus likely that 
the equality
$\chi = \chi_0 = \chi_\rho$
holds even at the nonlinear level.

Given that the structure of Eqs.~\eqref{eq_nonlin_closed} is similar to that found in TT theory,
we follow \cite{toner2012reanalysis} and 
conjecture that the scaling of correlation functions 
in the nonlinear theory is obtained using renormalized noise coefficients 
$\tilde{\Delta} = q_\perp^{z - \zeta} f_{\Delta}( q_\|/q_\perp^\xi)$ and renormalized dampings
$\tilde{\varepsilon} = q_\perp^z f_\varepsilon(q_\|/q_\perp^\xi)$
(where we have defined 
$\zeta\equiv d-1 + 2\chi + \xi$),
with functions $f_{\Delta}$ and $f_\varepsilon$ expected to be universal and to satisfy
$ f_{\Delta,\varepsilon}(x) \xrightarrow[x\to 0]{}{\rm cst.}$, 
$f_\Delta(x) \xrightarrow[x\to\infty]{} x^{(z - \zeta)/\xi}$, and 
$f_\varepsilon(x) \xrightarrow[x\to\infty]{} x^{z/\xi}$. 
On the other hand the speeds $c(\theta_{\bm q})$ should not be renormalized. 
Under all these assumptions, it is possible to predict the asymptotic behavior of equal-time correlation functions. 
For instance:
\begin{equation}
\langle |\delta\hat{g}_\perp(q_\|) |^2 \rangle  \underset{q_\| \to 0}{\sim} q_\|^{-\zeta/\xi},
\;\;\;
\langle |\delta\hat{g}_\perp(q_\perp) |^2 \rangle  \underset{q_\perp \to 0}{\sim} q_\perp^{-\zeta}.
\end{equation}
(For other functions, see \cite{SUPP}.)

\begin{figure}[t]
\includegraphics[width=\columnwidth]{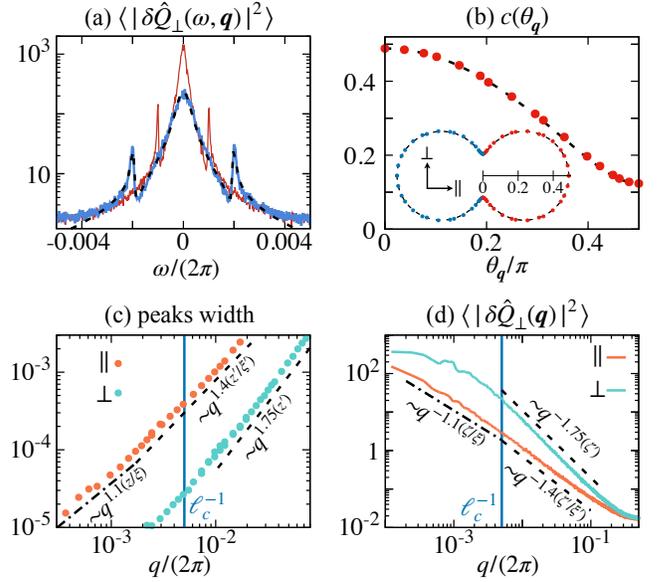}
\caption{Space-time correlations of fluctuations in the $L\ll \ell_{\rm r}$ regime ($\eta=0.1$, $L=8192$).
(a) Frequency spectra of order fluctuations at angle $\theta_{\bm q}=\tfrac{\pi}{4}$ 
with $\tfrac{q}{2\pi} = 0.002$ (red) and $0.004$ (blue).
The black dashed line is a fit by the theoretical predictions of Eqs.~\eqref{eq_space_time_cfs}.
(b) Angular dependence of the measured (dots) and predicted (dashed line) sound speed $c(\theta_{\bm q})$;
Inset: polar plot showing the $\pi$-symmetry of $c(\theta_{\bm q})$.
(c,d): scaling vs $q$ in the $\|$ and $\perp$ directions of frequency peak widths (c)
and equal-time order correlation function (d).
}
\label{fig2}
\end{figure}

We now come back to our Vicsek-style model at noise strength $\eta=0.1$
and show data for the order correlations confirming the structure of the above theory and providing estimates of the scaling exponents.
Additional results for the densities $\rho$ and $g_0$ will be published elsewhere~\cite{follow_up_paper}.
We actually measure the transverse nematic order $\delta Q_\perp$, which, when aligned along the horizontal direction 
and assuming small angular deviations, is a good proxy of $\delta g_\perp$.
($\delta Q_\perp \sim \cos(\theta)\sin(\theta) \sim \delta\theta\sim\bar{\rho}^{-1}\delta g_\perp$).

The frequency spectra do have the qualitative structure predicted by 
Eqs.~\eqref{eq_space_time_cfs}: two symmetric propagative peaks and a central diffusive one (Fig.~\ref{fig2}(a)).
As expected, peak locations, at a fixed angle $\theta_{\bm q}$ are proportional to $q$, allowing
the easy measurement of the sound speed $c(\theta_{\bm q})$, which we find in perfect quantitative 
agreement with Eq.~\eqref{eq_c} (Fig.~\ref{fig2}(b)). 
Peak widths provide estimates of $z$ and $z/\xi$ in the $\perp$ and $\|$ directions, as in TT theory. 
As shown in Fig.~\ref{fig2}(c), we find a crossover at the same scale $\ell_{\rm c}$ as observed in Fig.~\ref{fig1}(c). 
For scales below $\ell_{\rm c}$ we find $z'\simeq1.75$ and $z'/\xi'\simeq1.4$, 
while we are only able to estimate $z/\xi\simeq 1.1$ in the asymptotic regime 
(we use primes to denote exponent values measured below $\ell_{\rm c}$).
The equal-time order correlation function shown in Fig.~\ref{fig2}(c) in the $\|$ and $\perp$ directions,
also exhibits a crossover at $\ell_{\rm c}$.
From the pre-crossover scaling we estimate $\zeta'\simeq1.75$ and $\zeta'/\xi'\simeq1.4$, while we find  
$\zeta/\xi\simeq 1.1$ in the $q<2\pi/\ell_{\rm c}$ regime.

We thus have two sets of scaling exponents: for scales below $\ell_{\rm c}$, the above estimates lead to  $z'=\zeta'\simeq1.4$, $\xi'\simeq1.25$, and $2\chi'\simeq-0.5$. Note that this yields $-2\chi'/\xi'\simeq 0.4$, in agreement with our estimate of $\varpi'\simeq0.45$ in Fig.~\ref{fig1}(c) 
\footnote{A simple argument shows that $\varpi=-2\chi/\xi$, see \cite{SUPP}}. 
For scales beyond $\ell_{\rm c}$, we have $z=\zeta$, but cannot estimate $\xi$ from correlation functions.
Using $\varpi\simeq0.8$ (Fig.~\ref{fig1}(c)), yields $\xi\simeq1.1$ and $2\chi\simeq-0.9$, 
and finally $z=\zeta\simeq1.2$.
A few remarks are in order: (i) both below and above $\ell_{\rm c}$, $z=\zeta=1+2\chi+\xi$, a hyperscaling relation
also verified by polar flocks that implies that the dominant noises are additive and their amplitude is not renormalized
\cite{mahault2019TT}; (ii) in our nematic phase the anisotropy exponent $\xi\simeq1.1>1$, at odds with 2D polar flocks for which $\xi\simeq0.95<1$ \cite{mahault2019TT}, but in both cases we cannot exclude that scaling is asymptotically isotropic. 

To summarize, the orientational order emerging from self-propelled polar particles aligning nematically
 is always quasi-long-range asymptotically, but this regime is only observed
beyond $\ell_{\rm r}$, the scale associated to induced velocity reversals,
which can easily take very large values and often cannot even be measured.
Below $\ell_{\rm r}$, nematic order is fully long-range.
Constructing a hydrodynamic theory from microscopic grounds, 
we showed that this de facto phase has a structure and symmetries distinct from both conventional descriptions of active nematics and Toner and Tu theory.
Consequently, systems in the corresponding class exhibit features never reported so far, 
such as long-range nematic order and the presence $\pi$-symmetric propagative sound modes.

Finally, we believe our findings can be observed experimentally, 
as long as the rate of velocity reversals, be they induced or spontaneous, is small. 
After all, nematic alignment resulting from 
inelastic collisions between elongated objects is quite generic. Confined bacteria and motility assays are promising systems in this regard.

\acknowledgments
We thank Xia-qing Shi and Alexandre Solon for a critical reading of this manuscript. We acknowledge generous allocations of cpu time on the Living  Matter  Department  cluster  in  MPIDS,  and on  Beijing CSRC’s Tianhe supercomputer.

\bibliography{./Biblio-current.bib}

\end{document}